\documentclass[aip,reprint]{revtex4-1}

\draft 

\usepackage{graphicx}
\usepackage{amsmath}
\usepackage{float}
\usepackage{color}

\begin{document}

\title{First principles reactive simulation for equation of state prediction} 

\author{Ryan B. Jadrich}
\email{rjadrich@lanl.gov}
\affiliation{Theoretical Division, Los Alamos National Laboratory, NM, 87545}
\affiliation{Center for Nonlinear Studies, Los Alamos National Laboratory, NM, 87545}
\author{Christopher Ticknor}
\affiliation{Theoretical Division, Los Alamos National Laboratory, NM, 87545}
\author{Jeffery A. Leiding}
\email{jal@lanl.gov}
\affiliation{Theoretical Division, Los Alamos National Laboratory, NM, 87545}

\date{\today}

\begin{abstract}
The high cost of density functional theory has hitherto limited the \emph{ab initio} prediction of equation of state (EOS). In this article, we employ a combination of large scale computing, advanced simulation techniques, and smart data science strategies to provide an unprecedented, \emph{ab initio} performance analysis of the high explosive pentaerythritol tetranitrate (PETN). Comparison to both experiment and thermochemical predictions reveals important quantitative limitations of DFT for EOS prediction, and thus the assessment of high explosives. In particular, we find DFT predicts the energy of PETN detonation products to be systematically too high relative to the unreacted neat crystalline material, resulting in an underprediction of the detonation velocity, pressure, and temperature at the Chapman-Jouguet (CJ) state. The energetic bias can be partially accounted for by high-level electronic structure calculations of the product molecules. We also demonstrate a modeling strategy for mapping chemical composition across a wide parameter space with limited numerical data, the results of which suggest additional molecular species to consider in thermochemical modeling.
\end{abstract}


\maketitle 

\section{Introduction}
\label{introduction}

\begin{figure}
\centering
\includegraphics[width=3.37in,keepaspectratio]{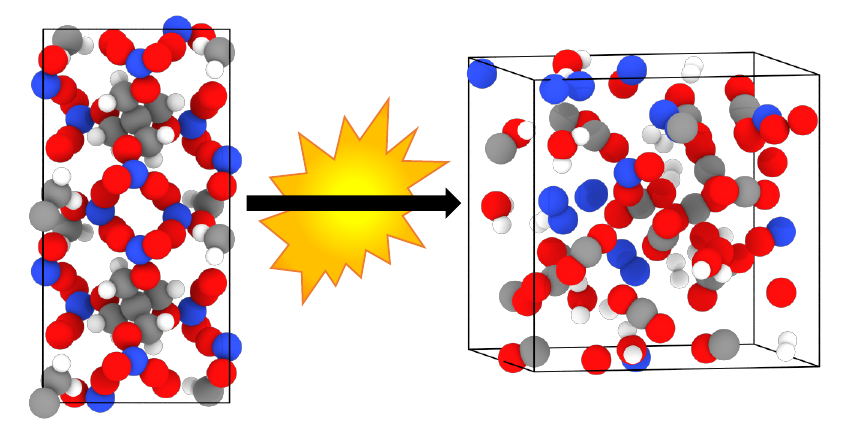}
\caption{Qualitative graphic of the atomic configuration changes that occur upon detonation of a typical molecular crystalline HE (PETN in this case) to that of a supercritical fluid of small molecule products (ex., carbon dioxide, water, molecular nitrogen).}
\label{fig:detonation}
\end{figure}

Predictive modeling of high explosive (HE) materials is a complex multi-scale physics problem whereby atomistic physics informs mesoscale modeling, enabling the prediction of macroscopic explosive performance properties (e.g., the detonation velocity and explosive work)~\cite{CooperBook,ZukasBook,ExplosiveOverview1,ExplosiveOverview2,ExplosiveOverview3}. Bridging the atomistic and mescoscopic scales is the HE equation of state (EOS), which is frequently decomposed into an EOS for the unreacted material and for the detonation products.~\cite{CooperBook,ZukasBook,ExplosiveOverview1,ExplosiveOverview2,ExplosiveOverview3} The former state is typically an organic crystal or liquid at standard conditions bound by dispersion forces.~\cite{DFTcrystals1,DFTcrystals2,DFTcrystals3} The latter state is generally a supercritical fluid of small molecules (e.g., $\text{H}_{2}\text{O}$, $\text{C}\text{O}$, $\text{C}\text{O}_{2}$, $\text{N}_{2}$).~\cite{CooperBook,ZukasBook,ExplosiveOverview1,ExplosiveOverview2,ExplosiveOverview3} A qualitative graphic depicting the detonation of pentaerythritol tetranitrate (PETN) from its molecular crystalline form to supercritical products is shown in Fig.~\ref{fig:detonation}. 

Predicting the thermodynamic properties for both the reactants and products requires condensed phase simulations. \emph{Ab initio} molecular simulation is the current state of the art in this respect. In terms of the trade-off between computational feasibility and accuracy, Kohn-Sham Density Functional Theory~\cite{KohnShamDFT,ElectronicStructureBook} (denoted as DFT in this work) is a popular choice for the calculation of forces and energies in these atomistic simulations. 


It is relatively straightforward to apply DFT to the reactant HE because the relevant range of thermodynamic conditions is often relatively limited (by comparison to product states) and chemical reactions are not occurring. Many studies have shown that DFT has great predictive capability.~\cite{DFTcrystals1,DFTcrystals2,DFTcrystals3,PETNVibrationCell}
Inclusion of dispersion interactions is generally necessary for condensed phase calculations to properly capture the cohesive energy. Post-hoc DFT corrective interactions, like the Grimme D2 and D3 correction schemes,~\cite{Grimme,GrimmeOrganicCrystals} are sufficient.~\cite{DFTcrystals1,DFTcrystals2,DFTcrystals3,PETNVibrationCell} Successes of dispersion corrected DFT (often termed DFT-D) include the accurate prediction of crystal cell parameters (and thus density), equation of state, and even vibrational modes.~\cite{DFTcrystals1,DFTcrystals2,DFTcrystals3,PETNVibrationCell} Accurate prediction of vibrational modes (phonons),~\cite{SolidStatePhysics,QMVibEffects} is critical to accounting for the zero-point energy~\cite{SolidStatePhysics,QMVibEffects} missed in a standard \emph{ab initio} simulation where nuclei are propagated classically.~\cite{DFTcrystals1,DFTcrystals2,DFTcrystals3,PETNVibrationCell,QMVibEffects} Quantum corrections can be important near the standard state ~(298 K and 1 bar), which unreacted HEs often are. The overall success of DFT for characterizing unreacted HEs motivates its application to study detonation.

Simulation of the products requires vastly more computational time to chemically equilibrate small molecule products at a vast array of potentially relevant state points. Roughly speaking, detonation conditions can span temperatures from 1,000-10,000 K and densities from 0.8 to 2.5 g/cc.~\cite{ExplosiveOverview1,ExplosiveOverview2,ExplosiveOverview3} Within these ranges, many state points will require long simulation times to establish chemical equilibrium, which is a slow activated process dominated by the breaking and re-forming of strong chemical bonds.~\cite{ShockMethane,Haber,NitrogenOxygenMix} High density is also likely to complicate matters by slowing the diffusion of atoms by geometric crowding. 


To circumvent the computational challenges of a comprehensive detonation products study, researchers have successfully leveraged alternative (more approximate) methods than DFT. The well known ReaxFF force field is a popular choice among such computationally expedient approaches.
ReaxFF leverages a many-body, bond-order type interaction to furnish a fully reactive potential.~\cite{ReaxFFOriginal,ReaxFFReview} Large simulation sizes and long run times are easily achieved. Studied materials include an impressive array of high explosives.~\cite{ReaxCarbonCluster,ReaxHECJ,ReaxQMDNewHE,ReaxQMDSilicatedPETN,ReaxShockPETN,QMVibEffects,ReaxNM} While successful, it is reasonable to assume that classical models will have a more restricted range of applicability than DFT given the coarse-grained electronic degrees of freedom.~\cite{GrimmeOrganicCrystals}.

Modern machine learning (ML) force fields are providing additional options for computationally tractable chemically reactive simulations.~\cite{Chimes1,Chimes2,Chimes3,Chimes4,Chimes5,ANI1,ANI2,ANI3,SNAP1,SNAP2,SNAP3,Rampi1,Rampi2,DFTB}. ChiMES is one such approach that has been applied to chemically reactive fluids. Successful applications include the modeling of carbon clustering in C/O systems at extreme conditions~\cite{Chimes4} and the detonation state of $\text{HN}_{3}$.~\cite{Chimes3} A major advantage of ML potentials are their ability to rapidly iterate and train on new data in a mostly automated fashion, a power that we leverage in this work. Like ReaxFF though, these models also integrate over electronic degrees of freedom to produce an effective classical potential. 

The very diverse range of conditions relevant to detonation products coupled with a desire to perform maximally predictive and minimally parameterized simulations, strongly motivates the use of finite-temperature DFT based approaches.\cite{ThermalSmearing1, ThermalSmearing2} For example, thermal effects on the electronic populations (excited states) are naturally accounted for in finite-temperature DFT by modification of the functional to include the electronic entropy.~\cite{ThermalSmearing1,ThermalSmearing2} Accounting for thermal electronic excitations is certainly important at higher temperatures as, in general, excitations will lead to a general softening of bonds and increased reactivity, more so than would be expected from enhanced kinetic energy of the atoms alone. High densities are likely to yield complications too. The very notion of well-defined molecular entities is challenged at highly compressed thermodynamic states, often of relevance to shocks or detonations. Extreme conditions can lead to non-standard behavior, such as the formation of electrically conductive states, which often possess unique extended structures~\cite{ShockMethane,FClConduction,ConductiveIce}. Classical force fields that ``integrate over'' the electronic degrees of freedom are unlikely to be predictive at such state points unless they were specifically calibrated for these regions.

In this paper we employ novel simulation and data-analysis techniques to overcome the computational burden of applying DFT to HE products. We employ a Monte Carlo (MC) simulation scheme that uses a machine-learned (ML) reactive force field, akin to those described above, as a reference potential to accelerate the simulation of reactive atomic systems while exactly retaining DFT level results. The ML potential is used in a nested Monte Carlo (NMC)~\cite{HeAr,Gelb,Iftimie,ShockNitrogen,ShockArgon,HydrogenFlouride,MethanolMethane,LayeredNMC} framework to propose a long chain of small moves that are then accepted or rejected \emph{in toto} according to how well the ML potential predicts the energy change. This scheme avoids having to compute the expensive DFT potential for every small update. In addition, we fit noisy simulation data via a machine learning based EOS that (1) enforces thermodynamic consistency and (2) simultaneously utilizes energy and pressure measurements to fit the model. Machine learning is also leveraged to maximize the information available in chemical composition data.

In addition to getting unprecedented accuracy for the product states, reactive \emph{ab initio} detonation products simulations are uniquely suited to validating and improving the thermochemical models of high explosives. Thermochemical techniques leverage approximate statistical mechanical theories to model chemical equilibrium in a coarse grained, computationally expedient manner. Two common thermochemical codes include \texttt{Cheetah}~\cite{CheetahBook}, of Lawrence Livermore National Laboratory, and \texttt{Magpie}~\cite{Magpie, MagpieVal}, of Los Alamos National Laboratory. A key limitation of both aforementioned models is the need to \emph{a priori} specify a list of possible molecular products, their classical interaction potentials, and internal partition functions. First-principles reactive simulations are uniquely suited to test the assumptions regarding which molecular species are relevant to detonation performance calculations, as well as the resultant EOS. Fully accounting for the most relevant species in a thermochemical modeling approach is integral to getting the right answer for the right reasons; only then can one have some faith in the reliability of predictions coming from thermochemcial models at new, untested conditions.

Before outlining the manuscript, we briefly discuss the formation of solid carbon in HE detonation products. The formation of solid carbon (soot) in the detonation products of carbon-rich/oxygen-poor HEs has been the subject of much previous research\cite{shaw87,shawids,vanThiel87,viecelli99} including relatively recent work.\cite{watkins2017,Gustavsen17} Solid carbon (as diamond or graphite for example) is often included in the list of possible products of thermochemical codes like \texttt{Magpie}; however, PETN is a relatively oxygen-rich HE, and thus solid carbon is a minor consideration.\cite{Ree_petn} Indeed PETN has been accurately modeled without the inclusion of solid carbon.\cite{bourasseau} In this work, we also neglect the formation of solid carbon in the detonation products calculations via \texttt{Magpie} and obtain good agreement with experiment. Indeed, modeling solid carbon in a first-principles context would be very difficult, as one would need to simulate with very large simulation cells to model the phase separation directly, or perform a Gibbs-ensemble simulation where the solid phases are modeled with separate cells; however, the former is becoming within reach with the advent of accurate machine-learning interaction potentials.\cite{armstrong20,lindsey20} This difficulty motivated the choice of PETN as our first real-world HE application of the new algorithms outlined herein.

In this article, we discuss the results of a fully \emph{ab initio} simulation study of the high explosive pentaerythritol tetranitrate (PETN) covering an unprecedented range of thermodynamic conditions. DFT simulation is used to construct a products EOS, from which detonation properties are calculated and compared to experimental and thermochemical results spanning a range of initial unreacted HE densities. The combination of accelerated simulations and data-driven analysis, provides an unprecedented, \emph{ab initio} EOS for high explosive products. Sect.~\ref{methods} discusses the various ML methods for simulating, EOS construction, and modeling of molecular populations in addition to the electronic structure, simulation, and thermochemical modeling details. Sect.~\ref{results} presents our HE performance predictions in comparison to experiment and thermochemical modeling. We discuss the quantitative limitations of DFT and present a possible remedy. We conclude in Sect.~\ref{conclusions} and put forward the need for future studies of more high explosives to further assess thermochemical modeling and to test the generality of our DFT correction for quantitative predictions.

\section{Methods}
\label{methods}

\subsection{Density functional theory}
\label{dft}

Simulations and single-point calculations utilized Kohn-Sham DFT~\cite{KohnShamDFT,ElectronicStructureBook} with the BLYP~\cite{BLYP1,BLYP2} functional and Grimme D3 dispersion corrections~\cite{Grimme,GrimmeOrganicCrystals} (with the C9 contribution). We chose the BLYP functional for its generally good performance in regards to organic molecules.~\cite{} All calculations used the CP2K software package~\cite{CP2K1,CP2K2} with triple-$\zeta$ double polarization basis sets,~\cite{Basis} GTH pseudopotentials,~\cite{PseudoPotential1,PseudoPotential2} and were spin-unrestricted with a multiplicity of unity. Energy cutoffs for the plain wave and Gaussian contributions to the basis sets were 600 and 60 Ry respectively. Convergence was further aided by the NN50 density and derivative smoothing implemented in CP2K. With these choices, increasing the respective cutoffs to 1000 and 100 Ry yielded numerically insignificant differences for an array of representative configurations. Excited electronic effects were accounted for via Fermi-Dirac smearing;~\cite{ThermalSmearing1,ThermalSmearing2} as such, the ``energies'' output by CP2K are electronic free energies (and the forces are electronic potentials of mean force). Atomic configurations were sampled according to the electronic free energy to correctly account for the electronic contribution to the partition function. For equation of state calculations though, we need to access the total internal energy (not free energy), amounting to a simple removal of the electronic entropy contribution according to standard thermodynamics.

\subsection{Gas-phase electronic structure}
In Section \ref{detonation_predictions}, we found it necessary to estimate the errors of the relative energies of BLYP\cite{BLYP1, BLYP2} with respect to high-accuracy CCSD(T)\cite{RAGHAVACHARI_CCSD_T, HAMPEL_CCSD, Knowles_RCCSD, RCCSD_erratum, DEEGAN_CCSD_T} calculations. For these calculations, we optimized isolated molecules (unreacted PETN and several molecules found in the products) at the MP2/6-31G\cite{Moller_Plesset, AMOS_RMP2} level of theory and calculated single-point energies at the CCSD(T)/6-31G level of theory. All gas-phase electronic structure calculations were performed in the \texttt{MOLPRO}\cite{molpro_2020, MOLPRO-WIREs, MOLPRO} electronic structure package. Further details of how these energies were used are provided Section \ref{detonation_predictions}.

\subsection{Monte Carlo simulation}
\label{mc}

In this work, we perform Monte Carlo (MC)~\cite{Metropolis,Hastings,AllenTildesley,FrenkelSmit} simulations at the DFT(BLYP) level of theory as discussed above. MC, as opposed to molecular dynamics (MD),~\cite{AllenTildesley,FrenkelSmit} was chosen for this work so as to leverage nonphysical moves that (1) accelerate kinetically slow bond rearrangements and (2) execute large-scale translations and rotations of molecules. MC has the added benefit of sampling directly from the equilibrium distribution without approximation.~\cite{Metropolis,Hastings,AllenTildesley,FrenkelSmit} While standard MC implementation suffers from smaller moves/updates in between costly QM calculation than MD, often displacing only a handful of atoms at a time, we use the following strategies to circumvent this difficulty.~\cite{AllenTildesley,FrenkelSmit}

To mimic the all-atom updates afforded by MD in our MC simulations, we use force-biased moves.~\cite{ForceBias1,ForceBias2,ForceBias3} Force biasing uses both the magnitude and the direction of the QM forces to move every atom in the simulation. In particular, a displacement ($\delta x$) for each Cartesian direction is drawn from a probability $\propto \exp{[\beta F \delta x / 2]}$ with a maximum allowed displacement of $\delta x_{\text{max}}$ where $F$ is the corresponding Cartesian force component. The dominant factor governing the acceptance rate of atom displacements is the temperature, not the density, as the short and strong chemical bonds (not collisions between atoms of separate molecules) are the limiting factor. The dominant temperature dependence is captured via the empirical relation: $\delta x_{\text{max}} \equiv a + b\times T$ where $a=0.014 \ \text{\AA}$ and $b=6\times 10^{-6} \ \text{\AA}/\text{K}$

Two additional MC moves are useful in chemically reactive mixtures, namely, swap and cluster moves. Swap moves are straightforward to implement and involve picking two random atoms and attempting to interchange them.~\cite{SwapMoves,NitrogenOxygenMix,HeAr} The move is accepted or rejected according to the standard Metropolis acceptance criterion. Cluster moves, on the other hand, have greater flexibility in the exact implementation.~\cite{AllenTildesley,FrenkelSmit} In this work, we define a cluster as a group of atoms connected via the cutoff distance $d_{i,j} = c_{i} + c_{j} + \delta_{\text{B}}$ where $c_{i}$ is the covalent radius of atom $i$ and $\delta_{\text{B}}=0.2 \ \text{\AA}$. By randomly picking a ``seed'' atom, a cluster is recursively built and randomly displaced and rotated about the center of position (COP). The maximum allowed magnitude for a random displacement in each Cartesian direction takes into account the density via $\delta x_{\text{COP}}=0.31\times (1.0/\rho)^{1/3}$, which is the dominant factor in determining acceptance. Empirically, we found that setting the maximum allowed angular displacement in proportion to that of the positional displacement was effective: $\delta \alpha_{\text{COP}}=\kappa\delta x_{\text{COP}}$ where $\kappa\equiv 100 \ ^\circ/\text{\AA}$. Appendix~\ref{cluster_move} details the specific implementation and how these parameters affect the motion. 

Alone, swap and cluster moves only update a handful of atoms in between each costly QM calculation; to increase the efficiency we perform a long sequence of such moves via Nested MC (NMC).~\cite{Gelb,Iftimie,ShockNitrogen,ShockArgon,HydrogenFlouride,MethanolMethane,HeAr,LayeredNMC} Specifically, NMC simulates the system on an approximate reference potential ($U_0$) for a fixed number of moves. The ``full'' or DFT potential energy ($U$) is then calculated at the end points of the chain.
The whole chain of moves is accepted or rejected \emph{in toto} according to $\min [1, \exp[-\beta(\delta U - \delta U_{0})]]$. The acceptance probability is determined by the length of the nested chain, and the accuracy of the reference potential with respect to the full potential. To maximize the acceptance rate for the nested chain of moves, we leverage a machine learned (ML) many-body potential for $U_{0}$.~\cite{HeAr} Details regarding the ML potential are discussed in Appendix~\ref{reference_potential} along with the training details. We note that any ML potential~\cite{Chimes1,Chimes2,Chimes3,Chimes4,Chimes5,ANI1,ANI2,ANI3,SNAP1,SNAP2,SNAP3,Rampi1,Rampi2,DFTB} that can handle chemical reactivity can in principle be used instead, with the corresponding acceptance probability being quantitative measures of their agreement with DFT. Force-biased moves and nested moves are performed randomly in a 50/50 ratio. Within the nested chain, swap and cluster moves are randomly performed in a 50/50 ratio.

An important aspect of our NMC approach is that it preserves exact sampling at the target level of theory (DFT, in this work) despite the approximate nature of the underlying reference force field. NMC can be considered an on-the-fly statistical re-weighting of configurations generated by the reference potential.~\cite{Gelb,Iftimie,ShockNitrogen,ShockArgon,HydrogenFlouride,MethanolMethane,HeAr,LayeredNMC} A poor overlap of the reference potential distribution with the exact distribution will result in a poor acceptance rate ($\alpha$) for nested moves. In this manner, $\alpha$ provides a real-time metric of how well the reference potential ensemble replicates the exact ensemble. For our reactive simulations an $\alpha \approx 50\%$ is achievable for nested chains of $\approx 30$ steps, with some variation due to the temperature. For reference, this is roughly an order of magnitude lower than what was realizable in a He/Ar mixture~\cite{HeAr} where the reference potential could be made virtually exact. For our current reactive system, the reference is less accurate than our previous work, but this is to be expected given the much greater complexity of super-critical reactive mixtures. Overall, our reference potential is sufficient for fast NMC sampling.

In total, 225 separate simulations were run on an evenly spaced $15\times 15$ grid of $\rho$ and $T$ with respective ranges [0.7, 2.5] g/cc and [1000, 8000] K. Each simulation started from randomly generated configuration by inserting 20 carbons, 16 nitrogens, 48 oxygen and 32 hydrogens (in that order). An insertion attempt was rejected if the distance between the inserted atom and any other previously inserted atom was smaller than the corresponding pairs mean Van der Waals radius. Skipping 20 frames at a time, a total of six snapshots were extracted from the end of each simulation on which to calculate statistics. While six snapshots is insufficient for calculating physical quantities at any one state, by utilizing all 225 simulations to fit regularized models for equation of state and molecular population (see Sect.~\ref{eos} and Sect.~\ref{molfind}), we are able to maximize the information content. In this manner, predictions at a state-point leverage the information from not one, but a whole array of neighboring simulations, leading to more confident estimates of physical quantities than would be expected from an isolated simulation.

In addition to reducing estimation error by fitting physically constrained models, our ``scattershot'' simulation approach in concert with the enhanced rare-event reactive sampling (relative to MD) helps erase sampling biases that might be present in the simulations due to a rugged energy landscape as more relevant energy minima will be sampled. More simulations must be run, but each can be run for shorter lengths of time as only a small amount of data at apparent equilibrium needs to be used. The bias reduction from our scattershot approach is not unlike that achieved in ``bagging''~\cite{hastie} (as used in machine learning) whereby a ``community of models'' yields an overall less biased result. Regardless, we note that it is always possible that the limited run times and system sizes affordable in \emph{ab initio} simulations could yield a bias that our approach cannot fully deal with. If such a bias is present, it will not be included in the uncertainty estimates which only probe the estimation error--a fact that is true of virtually any uncertainty analysis. 


\subsection{Equation of state modeling}
\label{eos}

To maximize the information content from our simulations, we leverage energy and pressure data simultaneously to train a Helmholtz free energy model with physically reasonable behavior. Specifically, we break the Helmholtz free energy per volume ($a$) into ideal ($a_{\text{id}}$) and excess ($a_{\text{ex}}$) contributions
\begin{equation} \label{eq:helmholtz}
a \equiv a_{\text{id}} + a_{\text{ex}}
\end{equation}
where
\begin{equation} \label{eq:helmholtz_ideal}
a_{\text{id}} \equiv k_{\text{B}}T n \big[\ln(n) - 1 - 3\ln(k_{\text{B}}T)/2\big]
\end{equation}
and $n$ is the number of atoms per unit volume. In Eqn.~\ref{eq:helmholtz_ideal}, constant terms have been dropped for numerical fitting convenience--these have no bearing on the modeling and serve only to scale out the units. The ideal entropy of mixing is also not explicitly in Eqn.~\ref{eq:helmholtz_ideal} as the atomic composition is (1) fixed throughout this work and (2) is irrelevant to the thermodynamic quantities of interest. The energy per volume and pressure are obtained by the standard relations: $e=a-T(\partial a / \partial T)$ and $P=n (\partial a / \partial n) - a$. Flexibility is embedded in the excess free energy contribution, which is described in Appendix~\ref{eos_model}. Uncertainty estimates are also calculated via bootstrap re-sampling of the data and the fitting procedure (see Appendix~\ref{eos_model}). 

To assist in training the EOS model, a baseline reference energy was subtracted from the raw simulation data to yield more manageable numbers. We define this reference energy ($\widetilde{E}$) as the sum of the heats of formation ($H_{\text{F}}$) for isolated $\text{C}\text{O}_{2}$, $\text{H}_{2}$, $\text{N}_{2}$ and $\text{O}_{2}$ molecules at standard conditions (1 bar, 298.15 K) in PETN relative abundances: $\widetilde{E}=5H_{\text{F}}(\text{C}\text{O}_{2})+4H_{\text{F}}(\text{H}_{2})+2H_{\text{F}}(\text{N}_{2})+H_{\text{F}}(\text{O}_{2})$. This choice is convenient as accurate experimental data is available and analogous DFT calculations are inexpensive and simple to perform. Specifically, DFT predictions leveraged standard statistical mechanical results for molecules whereby uncoupled translation, rotation and vibration is assumed.~\cite{McQuarrie1,Mcquarrie2}; the only required computation is that of a cheap small-molecule DFT normal mode analysis to parameterize the vibration partition function. The values for experiment~\cite{ExpHOF} and simulation, respectively, are $\widetilde{E}_{\text{EXP}}=-6.223\;\text{kJ/g}$ and $\widetilde{E}_{\text{DFT}}=-2198.654\;\text{kJ/g}$. All energies in this work, whether derived from experiment or DFT, will have the corresponding reference energy subtracted out.

\subsection{Thermochemical equation of state}
\label{thermochem}

Thermochemical modeling is performed with the \texttt{Magpie} software package developed at Los Alamos national laboratory.~\cite{Magpie}  \texttt{Magpie} approximates the free energy for a mixture of molecules via a decomposition into intra- and inter-molecular contributions. The former is approximated by splitting into independent electronic, vibration, rotation and translation contributions that are described by standard statistical mechanical forms .~\cite{McQuarrie1,Mcquarrie2} The electronic and vibration contributions are parameterized via electronic structure and/or experimental results. Inter-molecular interactions between molecules of the same type are approximated by Ross perturbation theory~\cite{Ross} in which molecules are coarse-grained to radially isotropic exponential-6 pair potentials. The thermodynamics of mixtures are then obtained by ideally mixing the pure molecular fluids. This may seem to be a crude approximation; however, it is very accurate at the conditions of relevance to high explosives products. This has been explicitly shown for the high explosive considered in this work (PETN).~\cite{MagpieVal} Our choice of \texttt{Magpie} product molecules are provided in Table~\ref{tab:Magpie_products} of Appendix~\ref{Magpie_products}.

\subsection{Jump conditions and Chapman-Jouguet theory}
\label{jump_cond}

The detonation front in a high explosive is well approximated as a near discontinuous change in all thermodynamic variables (e.g., energy, pressure, temperature, density) as well as a sudden change from unreacted high explosive (before the front) to equilibrated small molecule products (behind the front).~\cite{CooperBook,ZukasBook,ExplosiveOverview1,ExplosiveOverview2,ExplosiveOverview3} The locus of possible states that the products can be in is readily determined from the Rankine-Hugoniot jump conditions for mass, momentum, and energy conservation
\begin{equation} \label{eq:mass_conservation}
\dfrac{\rho}{\rho_{0}}=\dfrac{D}{D-u}
\end{equation}
\begin{equation} \label{eq:momentum_conservation}
P=\rho_{0}uD
\end{equation}
\begin{equation} \label{eq:energy_conservation}
E-E_{0}=\dfrac{1}{2}(P+P_{0})\Bigg(\dfrac{1}{\rho_{0}} - \dfrac{1}{\rho}\Bigg)
\end{equation}
where $D$ and $u$ are the detonation and material velocities, respectively, $\rho$ is the density (mass per volume), $E$ is the energy per mass of material, $P$ is the pressure (force per area), and the naught indicates the initial unreacted material state. Chapman-Jouguet theory then states that the single unique point with minimum $D$ (i.e., the CJ state) corresponds to the steady state detonation.~\cite{CooperBook,ZukasBook,ExplosiveOverview1,ExplosiveOverview2,ExplosiveOverview3} This unique state point is demarcated by a subscript with ``CJ'' throughout the text.

Both a DFT and experimental estimate for $E_{0}$ are calculated in reference to the baseline $\widetilde{E}$. Using the experimental $H_{\text{F}}$ for crystalline PETN~\cite{PETN_HOF} we find $E_{0,\text{EXP}}=4.52 \;\text{kJ/g}$. The analogous DFT value is derived from a standard phonon analysis of crystalline PETN. Specifically, two PETN molecules were placed in a periodic simulation cell with experimentally determined relative arrangements. The cell was scaled to the experimental density of 1.778 g/cc observed at 298 K.~\cite{PETNCrystalStructure}. Atomic positions were relaxed by an energy minimization and a phonon analysis of the energy minimized configuration was performed, the results of which parameterized the standard harmonic statistical mechanical model~\cite{SolidStatePhysics,McQuarrie1,Mcquarrie2}. Again with the reference energy removed, we find $E_{0,\text{DFT}}=4.33 \;\text{kJ/g}$, which is close to the experimental estimate. As mentioned in the Introduction (Sect.~\ref{introduction}), previous DFT-D modeling of crystalline high explosives found near quantitative performance. For the remainder of the text, all results are shown using $E_{0,\text{EXP}}$, though the detonation results are visually indistinguishable for the two.

\subsection{Molecular populations}
\label{molfind}

To approximately extract molecular species in our \emph{ab initio} simulation, we developed a graph-isomorphism software tool to label atomic clusters according to their topology of atomic connections. Clusters of atoms are defined by cutoff distances between atomic species and graphs are constructed using atoms as nodes and bonds as edges (edges are unweighted and only imply the cuttoff criteria is met). The bond cutoff distance for all atom pairs was set to the sum of the covalent radii plus a an extra threshold of 0.2 \AA \ except for Oxygen-Nitrogen which was set to 0.4 \AA. 
Graphs were constructed using the NetworkX python package which provides easy graph manipulation, characterization, and comparison abilities.~\cite{NetworkX} Two clusters (molecules) are considered identical if their respective graphs are isomorphic. Our approach does not account for charge, and cannot distinguish between isomers or enantiomers, but it is sufficient for the purposes of this work. Furthermore, our approach does not take into account the relative stability of a ``bonded'' pair as measured by the (1) bond order or (2) time stability. The former requires a complex analysis beyond the scope of this work, and the latter is better suited to molecular dynamics~\cite{Meyer2015,Sakano2020} with real time. Our analysis provides a coarse-grained structural overview of the chemistry only.

\section{Results and Discussion}
\label{results}

\begin{figure}
\centering
\includegraphics[width=3.37in,keepaspectratio]{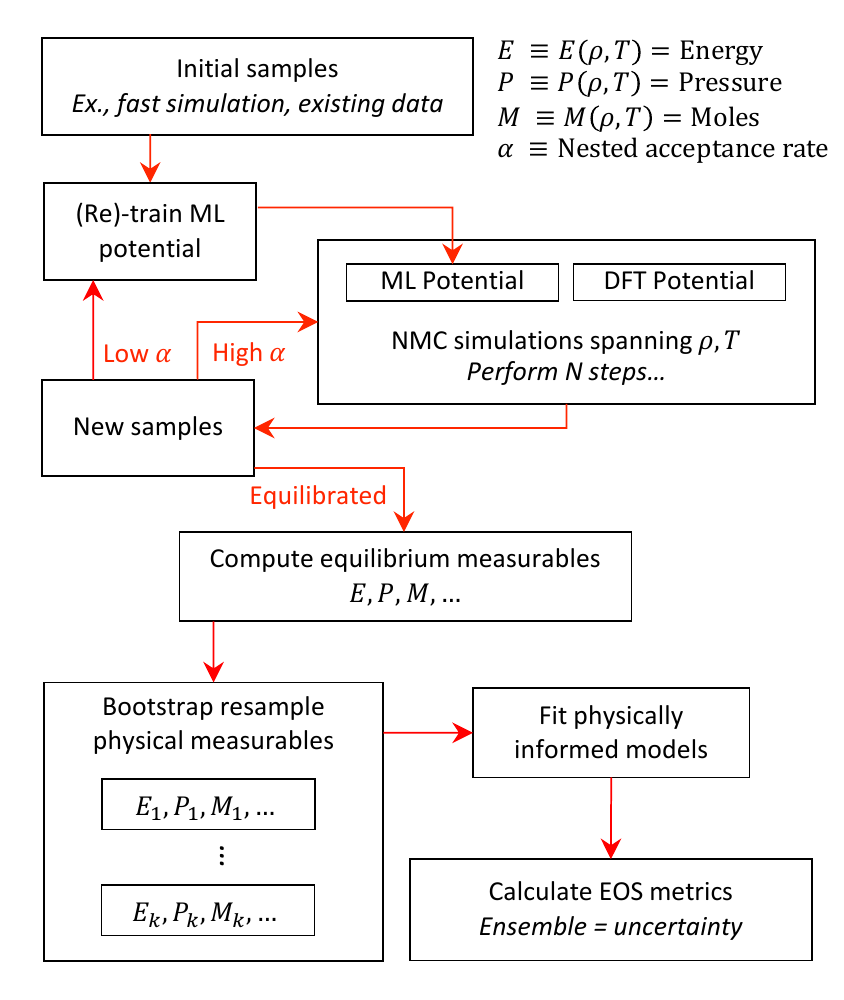}
\caption{The overall workflow employed in this study. Details about each stage/process are elaborated upon in Sect.~\ref{methods}}
\label{fig:workflow}
\end{figure}

By leveraging the advanced simulation and modeling efforts, outlined above, we ran 225 DFT simulations, covering a large range in state space--yielding a first principles EOS and some complementary molecular populations. Each simulation leveraged the ML enhanced NMC simulation approach of Sect.~\ref{mc}. Individual, noisy measurements from each simulation were not employed directly, but rather, were used to train our ML EOS model (Sect.~\ref{eos}). In doing so, we maximized the information content obtained from the many noisy simulation results and obtained a smooth, representation of the EOS with uncertainties. From our EOS, the performance of PETN is assessed over a large range of initial packing densities. Quantitative limitations of DFT simulation are found in respect to experiment. Similar to the EOS data, a model is trained to the noisy molecular composition data and a direct comparison of molecular populations predicted via DFT simulation and the thermochemical code \texttt{Magpie} is carried out for a limited set of species. Our results suggest the inclusion of another important molecular entity in \texttt{Magpie}. The overall workflow of our approach is summarized in Fig.~\ref{fig:workflow}. Specific details about each of the components can be found in Sect.~\ref{methods} and the relevant subsections.


\subsection{Detonation}
\label{detonation_predictions}

\begin{figure}
\centering
\includegraphics[width=3.37in,keepaspectratio]{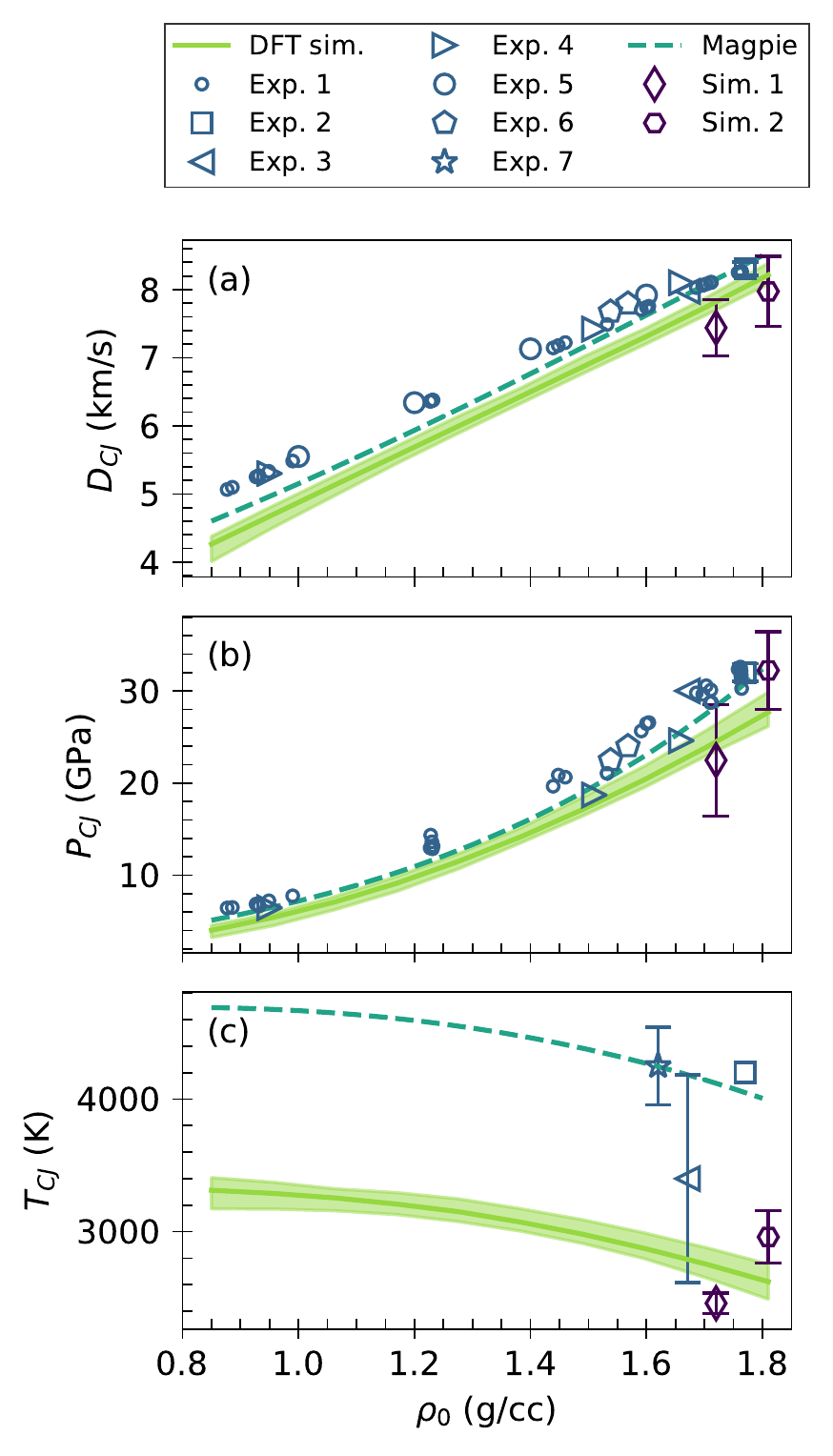}
\caption{Detonation velocity (a), Pressure (b) and Temperature (c) at the Chapman-Jouguet state. Uncertainty bounds are provided if available in the reference data sources. Thus, a lack of error bounds does not imply that the error is insignificant). Non-indexed entries correspond to this work whereas index values, Exp. 1~\cite{GreenLee}, 2-5~\cite{HornigLee}, 6-7~\cite{ColeburnChapman} and Sim. 1~\cite{ReaxHECJ}, 2~\cite{ReaxQMDSilicatedPETN}, come from the literature.}
\label{fig:CJ}
\end{figure}

As shown in Fig.~\ref{fig:CJ}a-b, our \emph{ab initio} (DFT simulation) predictions exhibit a systematic underprediction of $D_{\text{CJ}}$, and $P_{\text{CJ}}$ relative to a diverse set of experimental results. Fig.~\ref{fig:CJ}c suggests that $T_{\text{CJ}}$ may be underpredicted relative to experiment as well, but far less data are available and large (or unavailable) uncertainties obfuscate the comparison. It is unlikely that the differences between our \emph{ab initio} results and experiment can be attributed to experimental biases or uncertainties as the experimental measurements come from a variety of separate sources and are mutually consistent with one another in regards to $D_{\text{CJ}}$ and $P_{\text{CJ}}$. \texttt{Magpie} predictions are in better agreement with the experimental data, but again an overall underprediction is observed in $D_{\text{CJ}}$, and $P_{\text{CJ}}$. We note that the \texttt{Magpie} $T_{\text{CJ}}$ seems more reasonable, but the lack of definitive data makes it hard to come to any strong conclusions. 

In contrast to the experiments, our DFT simulation results are in reasonable agreement with two independent simulation studies, as shown in Fig~\ref{fig:CJ}. Both studies utilized the ReaxFF force field, however, only one refined the configurations with short \emph{ab initio} molecular dynamics runs. A similar degree of underprediction in $D_{\text{CJ}}$ to that found in this work was observed in each study. For $P_{\text{CJ}}$, the prior work brackets our prediction, though the uncertainties are fairly large. Furthermore, the prior simulation studies have CJ state temperatures that closely match our results, as seen in Fig.~\ref{fig:CJ}c. 

\begin{figure}
\centering
\includegraphics[width=3.37in,keepaspectratio]{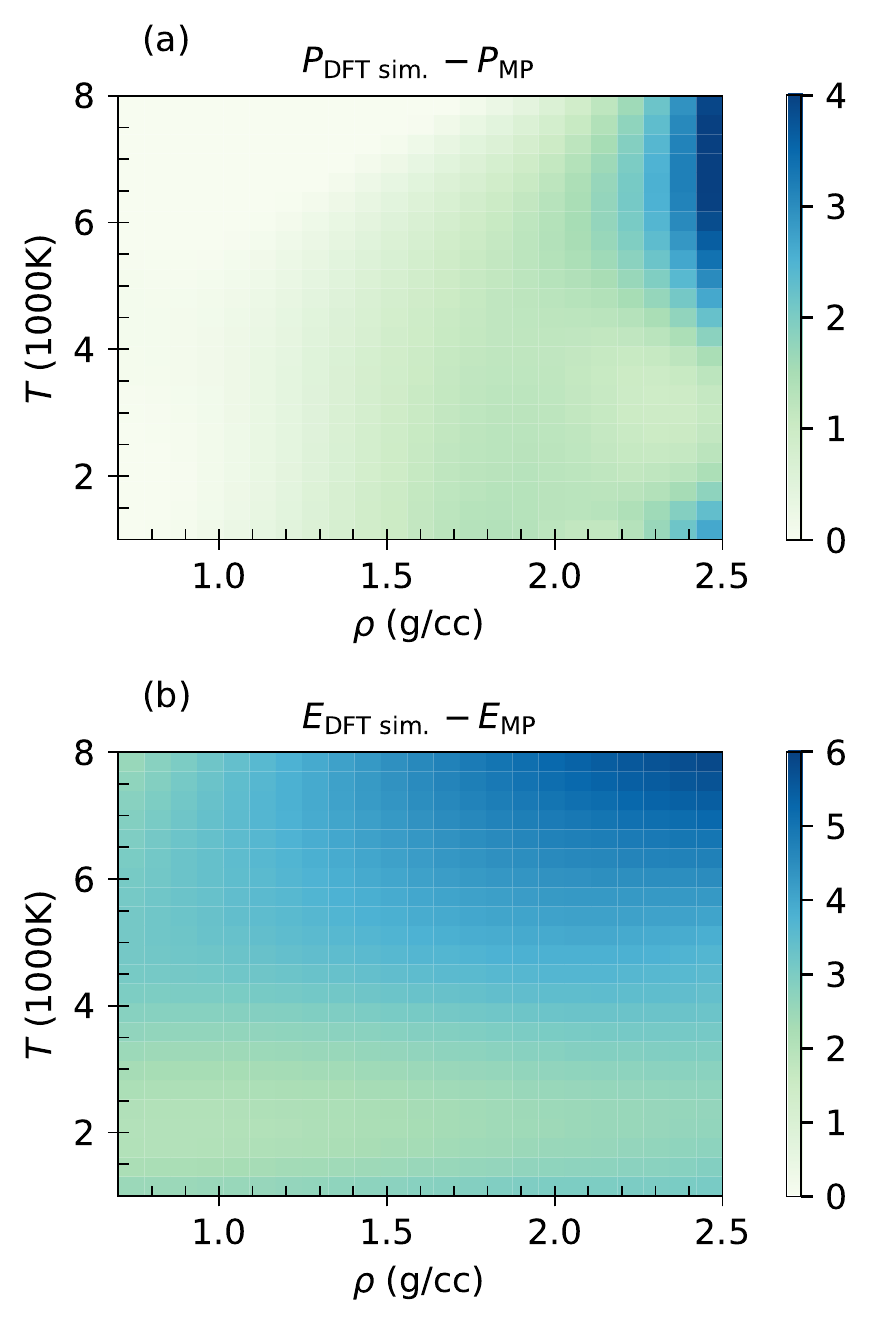}
\caption{Difference between the DFT simulation and \texttt{Magpie} pressure (a) and energy (b) predictions. The average of the DFT simulation results is over the uncertainty obtained by the model bootstrapping procedure outlined in Sect.~\ref{eos}}
\label{fig:ml_v_mp}
\end{figure}

Assuming no other complications related to chemical equilibrium or a breakdown of CJ theory, any discrepancies between experiment, theory and simulation must arise from differences in the underlying equilibrium equation of state. To help probe for issues in the DFT simulation predictions, it is useful to compare the underlying energy and pressure predictions between \texttt{Magpie} and simulation, especially since the former agrees better with experiment. As shown in Fig.~\ref{fig:ml_v_mp}a, the pressures from our DFT simulations and \texttt{Magpie} are in quite good agreement. Simulation tends to yield slightly higher pressures with increasing temperature, relative to \texttt{Magpie}, (can approach $\approx3$ GPa); however, both simulation and \texttt{Magpie} converge at lower temperatures and there appears to be no systematic bias. In comparison, the simulation energy is systematically elevated with respect to \texttt{Magpie} by $\geq 2 \ \ \text{kJ/g}$, as shown in Fig.~\ref{fig:ml_v_mp}b. The systematically higher DFT simulation product state energies certainly contribute to the systematic underprediction in $D_{\text{CJ}}$, $P_{\text{CJ}}$, and $T_{\text{CJ}}$ relative to \texttt{Magpie}. 

\begin{figure}
\centering
\includegraphics[width=3.37in,keepaspectratio]{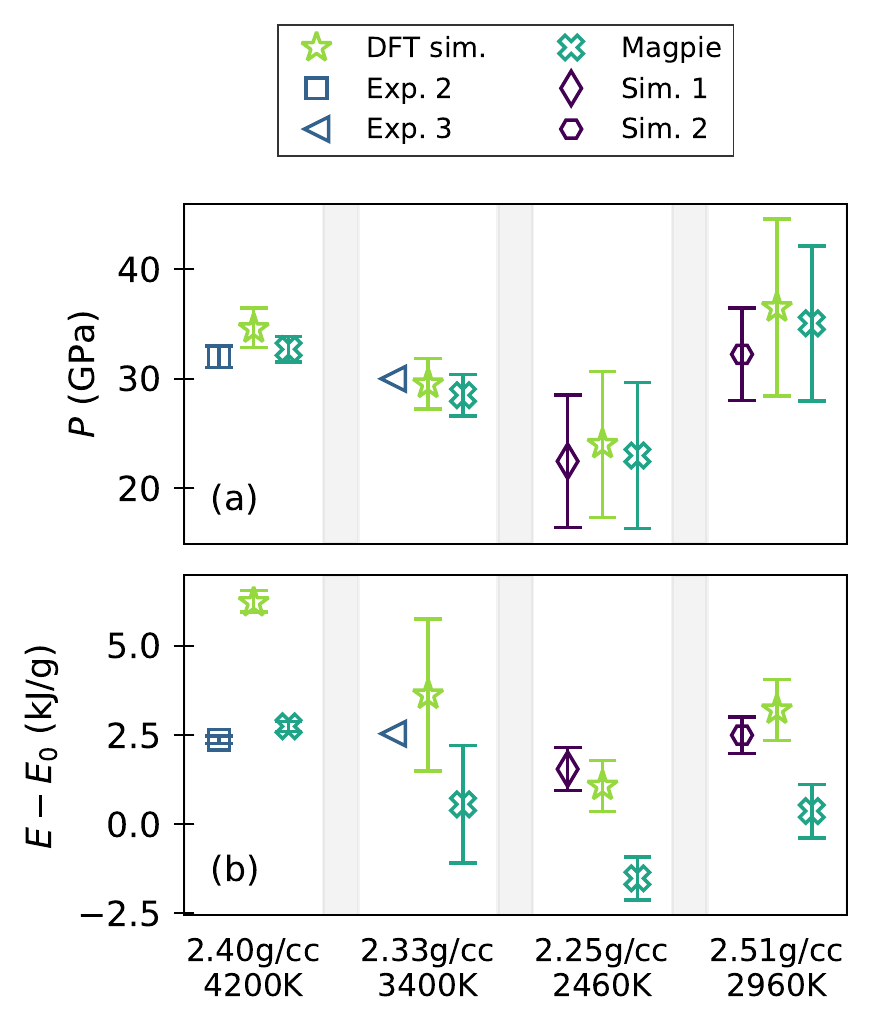}
\caption{
Comparison of DFT simulation and \texttt{Magpie} predictions at four state points for which either experimental or simulated pressure (a) or energy (b) measurements exist. The x-axis provides the state-points of relevance taken from the various literature results, the citations of which are provided in the figure legend. All uncertainty bounds are at the 95\% confidence interval. DFT simulation and \texttt{Magpie} uncertainties take into account any quoted uncertainty in the state points ($\rho, T$) obtained from the literature citations provided in the figure legend. A lack of uncertainty on literature data means only that it was not reported. Non-indexed entries correspond to this work whereas index values, Exp. 2-3~\cite{HornigLee} and Sim. 1~\cite{ReaxHECJ} and 2~\cite{ReaxQMDSilicatedPETN}, come from the literature.} 
\label{fig:E_P_comp}
\end{figure}

A comparison, similar to that in Fig.~\ref{fig:ml_v_mp}, can be carried out for select literature CJ state-points where energy, pressure, density and temperature are either all available or can be supplemented with the the jump conditions (Eqn.~\ref{eq:mass_conservation}-\ref{eq:energy_conservation}) to obtain any missing information. Experimental data with sufficient information are in limited supply due to the challenges with measuring a CJ temperature. While there is no difficulty in calculating properties from simulation, there are minimal data with which to compare. In total, we found four points with which compare our results, two simulations and two experiments. The four state-points are indicated on the x-axis of Fig.~\ref{fig:E_P_comp}, each with either a distinct experiment or simulation result from the literature; along with each are corresponding predictions from DFT simulation and \texttt{Magpie}. As shown in Fig.~\ref{fig:E_P_comp}a, the pressures are all in good agreement, though the uncertainties on some points are large. In contrast, the energies in Fig.~\ref{fig:E_P_comp}b generically disagree. First, DFT simulation energies are again seen to be systematically elevated relative to those of \texttt{Magpie}. Second, the two experimental results are in conflict: one agrees with \texttt{Magpie} and the other is in between DFT simulation and \texttt{Magpie}. Third, both previous simulations agree with our DFT simulation predictions as apposed to \texttt{Magpie}. Further experimental studies with CJ temperature measurements will be critical to resolving these disparities.

\begin{figure}
\centering
\includegraphics[width=3.37in,keepaspectratio]{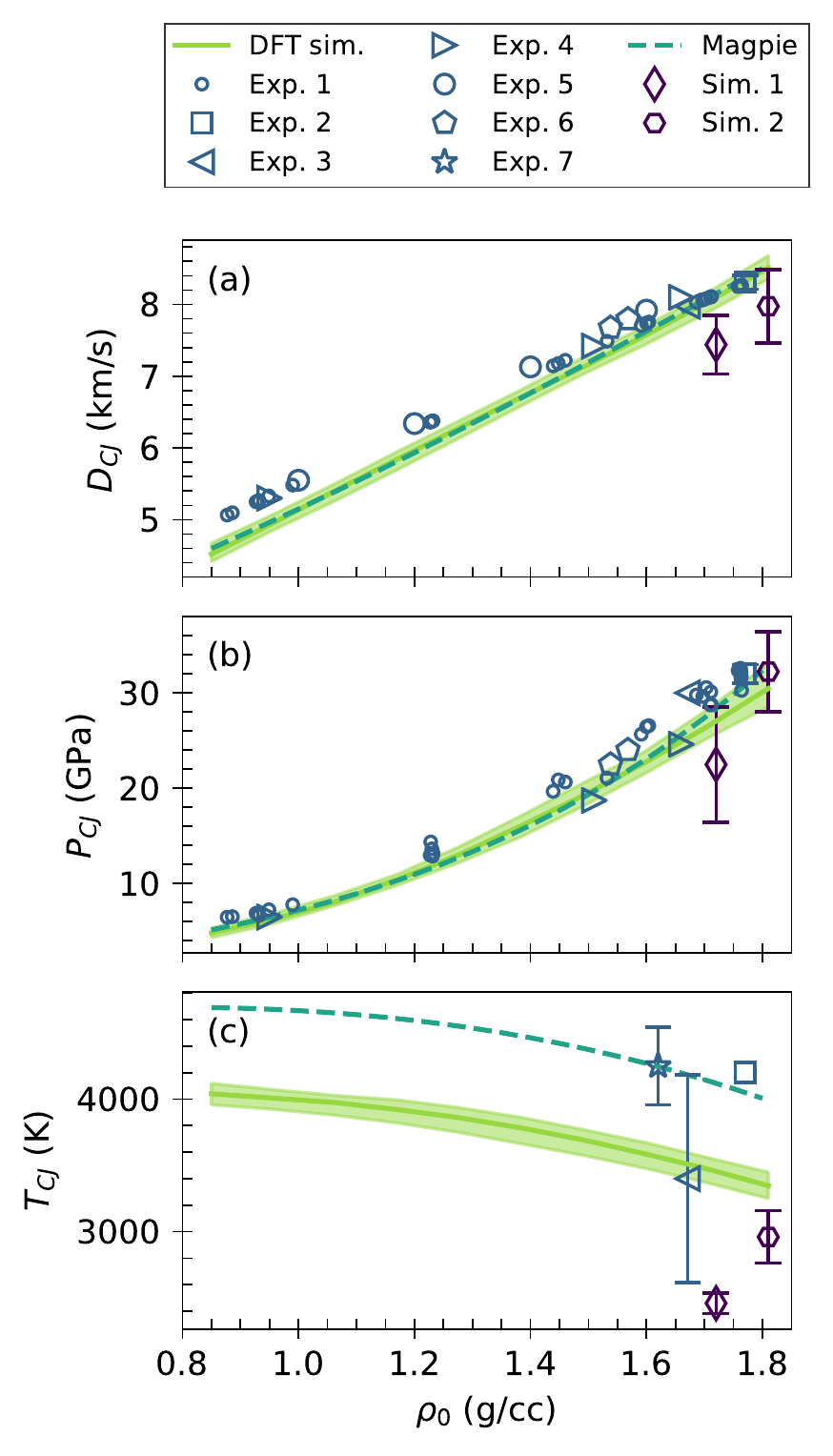}
\caption{Detonation velocity (a), Pressure (b) and Temperature (c) at the Chapman-Jouguet state. Uncertainty bounds are provided if available in the reference data sources. Thus, a lack of error bounds does not imply that the error is insignificant. Non-indexed entries correspond to this work whereas index values, Exp. 1~\cite{GreenLee}, 2-5~\cite{HornigLee}, 6-7~\cite{ColeburnChapman} and Sim. 1~\cite{ReaxHECJ}, 2~\cite{ReaxQMDSilicatedPETN}, come from the literature. (Same as Fig. \ref{fig:CJ} except a \emph{post hoc} energy correction as described in the text has been applied to the DFT simulation results.)}
\label{fig:D_P_T_CJ_shift}
\end{figure}

Application of a \emph{post hoc} energy correction to our DFT simulations, derived from high-level electronic structure calculations of the detonation products, is enough to bring our predictions into substantially better agreement with experiment. Specifically, CCSD(T) energy calculations of the detonation products may provide a full \emph{ab initio} alternative to calibrating the energy to known experimental measurements. We estimated an inherent error in the energy of the products due to the BLYP functional with the following procedure. We calculated $\Delta \Delta E = \left(E_{\text{p,BLYP}} - E_{\text{r,BLYP}}\right) - \left(E_{\text{p,CCSD(T)}} - E_{\text{r,CCSD(T)}}\right)$, which represents the shift in the energy difference between PETN in the product and reactant states of BLYP with respect to CCSD(T). Compositions for the products states are required to calculate $\Delta\Delta E$ at various densities and temperatures. We find that DFT compositional ambiguities, due to the sometimes vague notion of what constitutes a molecule in a reactive system, complicates the definition of a bulk molecular composition; thus, we choose the well-defined chemical compositions predicted by the \texttt{Magpie} thermochemical code. We restrict ourselves to the gas phase in calculating $\Delta \Delta E$ for obvious computational reasons. This is reasonable, as a large portion of the energy in a molecular fluid comes from the chemical bonding itself. Along the entire CJ locus we find a correction of about 1.8 kJ/g; 
this is 75 \% of the value shown in Figure \ref{fig:ml_v_mp} near the same locus of states, suggesting that the bulk of the inherent error in the BLYP energy of the CJ products can be determined by this recipe. Once we've calculated the $\Delta\Delta E\left(\rho, T\right)$ surface, we add this to the energy surface used in our CJ calculations. This produces a corrected CJ locus, shown in Figure \ref{fig:D_P_T_CJ_shift}, which is in better agreement with the experimental data and \texttt{Magpie} calculations. 

To be clear, we do not assert that this process will work as well every time and for every HE. We merely present it here as a possible recipe to approximate the error in one's choice of DFT functional for the condensed phase calculation. This, very importantly, allows the method to persist as a predictive capability, and thus, it remains free from requiring experimental data on which to calibrate an energy shift of the DFT-products energy surface. We are currently performing simulations of HMX, DAAF, and TATB. These future studies will help determine the generality of the energy correction procedure described here.

A second potential source of error is the lack of quantum-nuclear molecular vibrations; however, we explored this possibility by re-calculating the \texttt{Magpie} CJ locus in Figure \ref{fig:CJ} wherein the quantum harmonic oscillators were replaced with classical ones, and found very small differences in $U_s$ (0.25-0.5\%), $P$ (1-1.5\%) and $T$(1.5-1.8\%), implying that this is not a significant source of error. 

\subsection{Molecular composition}
\label{molecular_composition}

\begin{figure}
\centering
\includegraphics[width=3.37in,keepaspectratio]{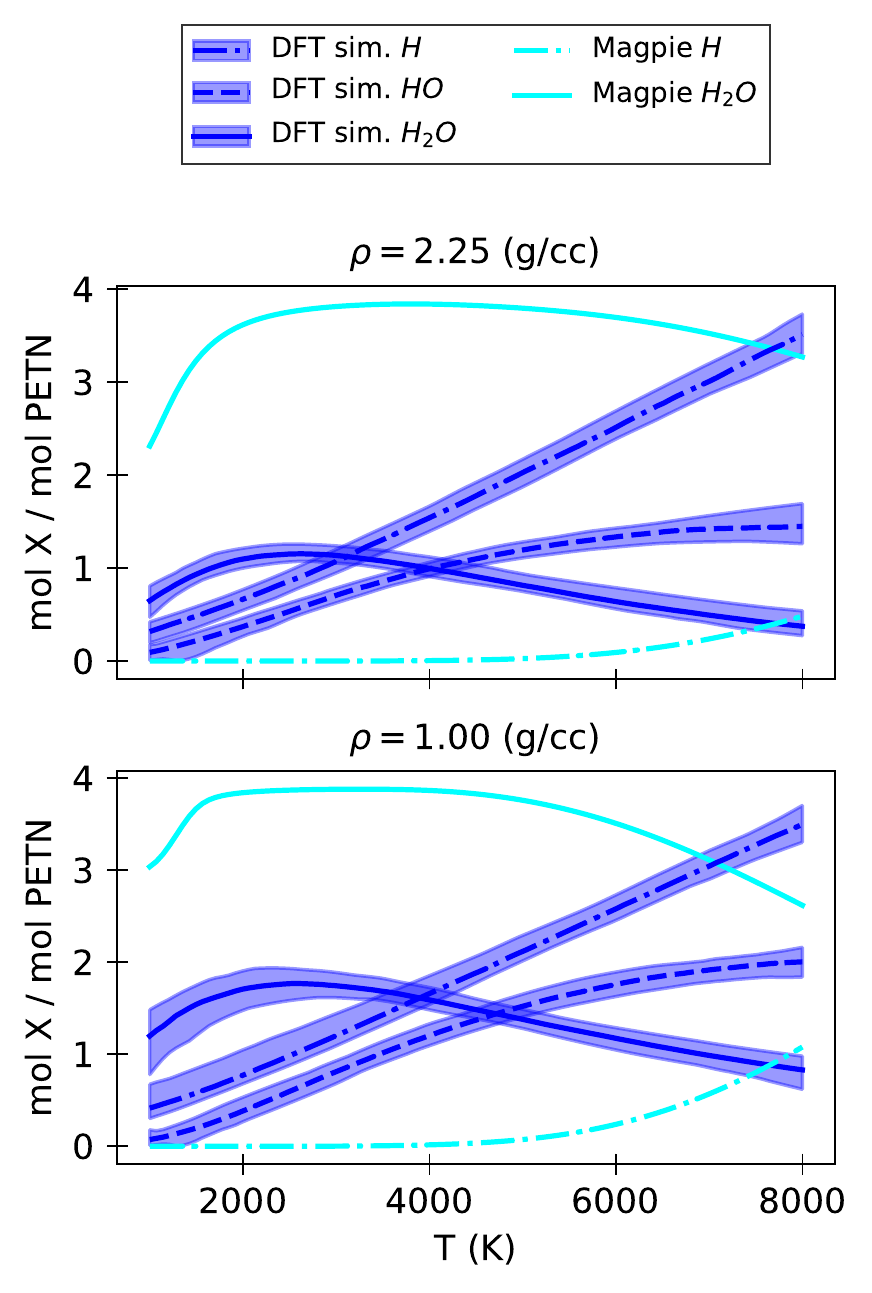}
\caption{Composition of water ($\text{H}_{2}\text{O}$), hydroxyl ($\text{HO}$), and atomic hydrogen ($\text{H}$) along two isochores as predicted by DFT simulation and \texttt{Magpie}.}
\label{fig:mol_comp_isc_H2O_HO_H}
\end{figure}

First-principles simulation provides a unique window into microscopic chemical details that can be used to assess and improve thermochemical modeling. As mentioned in Sect.~\ref{thermochem}, thermochemical models like \texttt{Magpie}~\cite{Magpie,MagpieVal} and \texttt{Cheetah}~\cite{CheetahBook} require a selection of possible product molecules. For high explosives composed of $\text{C}$, $\text{H}$, $\text{N}$, and $\text{O}$, typical entities are shown in Table~\ref{tab:Magpie_products}. The choice of species to include is generally motivated by many previous theoretical works,~\cite{FriedEXP62002, Charlet1998, Ree_petn, Ree_supercrit} as well as experiment~\cite{Ornellas}. Optimizing intermolecular parameters for new species is a non-trivial effort, as such, the prospect of narrowing the scope for new molecules via \emph{ab initio} simulation, is enticing, and will be addressed in future works.

\begin{figure}
\centering
\includegraphics[width=3.37in,keepaspectratio]{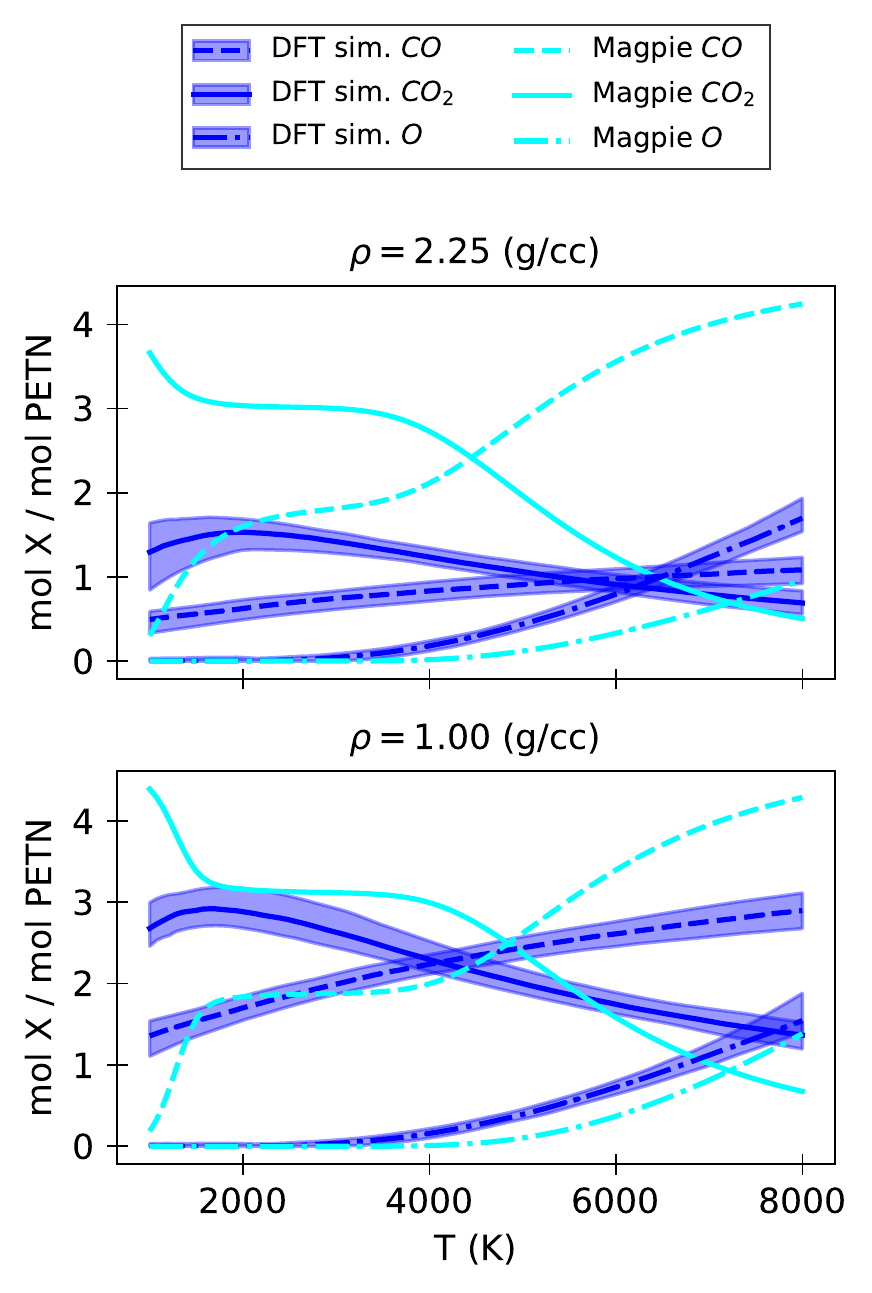}
\caption{Composition of carbon dioxide ($\text{CO}_{2}$), carbon monoxide ($\text{CO}$), and atomic oxygen ($\text{O}$) along two isochores as predicted by DFT simulation and \texttt{Magpie}.}
\label{fig:mol_comp_isc_CO2_CO_O}
\end{figure}

As demonstrated in Fig.~\ref{fig:mol_comp_isc_H2O_HO_H}, our DFT simulations strongly support the inclusion of $\text{HO}$ (hydroxyl) with $\text{H}$ and $\text{H}_{2}\text{O}$ for thermochemical modeling. The lack of $\text{HO}$ in \texttt{Magpie} is undoubtedly a major driving force behind the observed discrepancies in the populations of $\text{H}_{2}\text{O}$, and $\text{H}$. In general \texttt{Magpie} has much more water than the simulations and the inclusion of $\text{HO}$ would provide a mechanism for its reduction. While the presence of $\text{HO}$ has some precedence in the literature,~\cite{therm_chem_params_1,ReaxHECJ,ReaxQMDSilicatedPETN} much less is known regarding the general importance of including $\text{HO}$ for a typical HE. In contrast, Fig.~\ref{fig:mol_comp_isc_CO2_CO_O} shows a qualitative agreement between the simulation and \texttt{Magpie} populations of $\text{CO}_{2}$, $\text{CO}$, and $\text{O}$; the agreement is better at lower density where many body crowding effects are minimal and the ``mean field'' assumptions in \texttt{Magpie} are more accurate. The importance of other carbon/oxygen species is not totally ruled out by this analysis, but it is suggestive that the chemical description is at least qualitatively adequate for carbon-oxygen species. A more complete assessment of simulated molecular populations and how they compare to thermochemical codes necessities further simulation. Specifically, data on multiple high explosives will provide suggested improvements and molecular species for inclusion that are relevant across the board, and not just possibly for one HE, which is the ultimate goal.

\section{Conclusions}
\label{conclusions}

We have demonstrated a modern computational strategy that enables the \emph{ab initio} performance analysis of a high explosive over an unparalleled range of conditions. Crucial to the success of our approach is the combination of (1) an efficient Monte Carlo simulation framework catered to reactive systems and (2) a machine learning based equation of state model to maximize the information content available from the many noisy simulations. CJ state predictions for the detonation velocity, pressure and temperature were obtained over an unprecedented range of initial unreacted HE densities.

DFT(BLYP) simulation systematically under-predicts the CJ detonation velocity, pressure and temperature relative to both experiment and thermochemical predictions. This discrepancy is attributable to an overestimated energy gap between the product molecule statepoints and the unreacted crystalline high explosive. Interestingly, the energy discrepancy may be mostly accounted for by \emph{post hoc} high-level electronic structure calculations, CCSD(T), on the product molecules. Using product compositions obtained from \texttt{Magpie}, CCSD(T) calculations suggest that BLYP actually overestimates the energies of isolated product molecules--which in the relative abundances for PETN detonation products--roughly amounts to $\approx1.8$ kJ/g along the CJ locus. For a select set of DFT configurations where molecule definitions are clear-cut, similar energy corrections are found.

Important future extensions of this work are three-fold. First, new high explosives must be examined to see if the \emph{post hoc} energy correction, described above, is robust. If yes, this implies that a quantitatively accurate first-principles performance analysis of a HEs is within reach, and importantly, without a need for calibrating to a small number of experimental measurements. Work along these lines is already underway for HMX, DAAF, and TATB. Second, the product molecules produced by \emph{ab initio} DFT simulation warrant a closer examination for species of possible importance to thermochemical modeling. Such information has the potential to validate and improve thermochemical modeling efforts in a manner that is complementary to experiments. Key information beyond what we explored in this work includes the electronic environment of the molecular aggregates to differentiate between charged and radical species. Third, our approach can be used as a validation tool for ML force fields. Specifically, by using ML force fields directly, much greater system sizes and time scales are accessible. The reliability of these predictions can be probed by comparing the results of small-scale ML + NMC simulations with direct ML simulations. 

\begin{acknowledgments}
R.B.J. acknowledges funding from the Nicholas C. Metropolis Postdoctoral Fellowship and the ASC-PEM-HE program at Los Alamos National Laboratory. J.A.L. and C.T. acknowledge funding from the ASC-PEM-HE program at Los Alamos National Laboratory. This work was supported by the US Department of Energy through the Los Alamos National Laboratory. Los Alamos National Laboratory is operated by Triad National Security, LLC, for the National Nuclear Security Administration of U.S. Department of Energy (Contract No. 89233218NCA000001).
\end{acknowledgments}

\section*{Data availability statement}
\label{avail}
The data that support the findings of this study are available from the corresponding authors upon reasonable request.

\appendix

\section{Monte Carlo cluster moves}
\label{cluster_move}

A cluster move is a composed from the following sequence:
\begin{enumerate}
  \item Randomly pick a seed atom.
  \item Build a cluster from a set of atom-atom cutoff distances based on the atom types.
  \item Unwrap the coordinates from periodic imaging to ensure a contiguous representation of the cluster is obtained.
  \item Randomly displace the cluster center of position (COP).
  \item Randomly rotate the cluster about the COP.
  \item Reapply periodic boundary conditions to the cluster.
  \item Using the same seed atom, attempt to build the cluster again and reject the move if not possible due to the formation of new connections (required for preserving detailed balance).~\cite{AllenTildesley,FrenkelSmit}
  \item If not already rejected, perform a standard, unbiased, metropolis acceptance rejection check based on the potential energy change.~\cite{AllenTildesley,FrenkelSmit,Metropolis,Hastings}
\end{enumerate}
The COP displacement (Step 4) is of the standard form used in MC: each Cartesian component is displaced by a uniformly randomly sampled amount in the interval $[0,\delta x_{\text{max}}]$. The random rotation (Step 5) is more involved and elaborated on below.

A random 3D rotation about the COP is realized via successive rotation of the Euler angles $(\phi,\theta,\psi)$ to the COP removed Cartesian coordinate vector $\boldsymbol{x}$. The first and last rotations are about the z-axis, which we denote by the rotation matrix $\boldsymbol{\mathcal{R}}_{z}(\gamma)$ where $\gamma$ is the rotation angle. The middle rotation is about the x-axis and is denoted as $\boldsymbol{\mathcal{R}}_{x}(\gamma)$. The rotated coordinate is obtained via
\begin{equation} \label{eq:euler_rotation}
\tilde{\boldsymbol{x}} \equiv \boldsymbol{\mathcal{R}}_{z}(\psi) \boldsymbol{\mathcal{R}}_{x}(\theta) \boldsymbol{\mathcal{R}}_{z}(\phi) \boldsymbol{x}
\end{equation}
The reverse rotation is obtained by negating the angles and swapping $\psi$ and $\phi$, yielding
\begin{equation} \label{eq:euler_rotation_rev}
\boldsymbol{x} \equiv \boldsymbol{\mathcal{R}}_{z}(-\phi) \boldsymbol{\mathcal{R}}_{x}(-\theta) \boldsymbol{\mathcal{R}}_{z}(-\psi) \tilde{\boldsymbol{x}}
\end{equation}
The most straightforward way to ensure detailed balance (actually, super-detailed balance) is to guarantee that the reverse rotation is equally likely to be proposed as the forward rotation. We realize this by uniformly randomly sampling each angle from the symmetric range $[-\alpha_{\text{max}},\alpha_{\text{max}}]$ where $\alpha_{\text{max}}$ is the maximum allowed angular rotation mentioned in Sect.~\ref{mc}. Technically, $\theta$ can be sampled from a different symmetric range than $\phi$ and $\psi$, but setting all of them equal was effective and convenient.

\section{Machine learned reference potential}
\label{reference_potential}

Our machine learning model assumes a decomposition of the total system energy into individual atom energies (the locality assumption). Furthermore, the energy of an atom is decomposed into six unique contributions. Without specifying the exact functional forms yet, the first two contributions correspond to explicit two- and three- body correlations via
\begin{equation} \label{eqn:two_body}
E_{2}(i) \equiv \sum_{j}^{N} f_{2}(i,j)
\end{equation}
and
\begin{equation} \label{eqn:three_body}
E_{3}^{(\kappa)}(i) \equiv \sum_{\substack{j,k,\\ j\neq k}}^{N} f_{3}^{(\kappa)}(i,j,k) 
\end{equation}
where $f_{2}$ and $f_{3}^{(\kappa)}$ with $\kappa \in [1,2]$ are yet to be specified functions of two and three atoms (denoted by their indices) respectively. Analagous to the explicit forms above, three different ``infinite'' body forms are used as well
\begin{equation} \label{eqn:two_body_mf}
\widetilde{E}_{2}(i) \equiv \prod_{j}^{N} [1 + f_{2}(i,j)]
\end{equation}
and
\begin{equation} \label{eqn:three_body_mf_2}
\widetilde{E}_{3}^{(\kappa)}(i) \equiv \prod_{\substack{j,k,\\ j\neq k}}^{N} [1 + f_{3}^{(\kappa)}(i,j,k)]
\end{equation}
In practice, the ``infinite body'' forms never involve all of the atoms in a system as a radial cutoff (described below) is imposed.

As originally used by~\citet{HeAr}, the specific two and three body terms are built from a radial (R) function and an angular (A) function as
\begin{equation} \label{eqn:two_body_explicit}
f_{2}(i,j) \equiv f_{\text{R}}(r_{i,j}) 
\end{equation}
and
\begin{equation} \label{eqn:three_body_explicit}
f_{3}^{(\kappa)}(i,j,k) \equiv f_{\text{R}}(r_{i,j})f_{\text{R}}(r_{i,k})[\delta_{1,\kappa}+\delta_{2,\kappa}f_{\text{A}}(\theta_{i,j,k})] 
\end{equation}
where $r_{i,j}$ is the radial distance between particles $i$ and $j$, $\theta_{i,j,k}$ is the angle formed between a vector pointing from particle $i \rightarrow j$ and from particle $i \rightarrow k$. The radial component is
\begin{equation} \label{eqn:radial_func}
f_{\text{R}}(r) \equiv \sum_{i=1}A_{i}\dfrac{e^{-r/Z_{i}}}{r}\cos[k_{i}r-Q_{i}]
\end{equation}
where $\{A_{i},Z_{i},Q_{i}\}$ is the set of flexible parameters with associated wave-vectors $\{k_{i}\}$. The corresponding angular term is
\begin{equation} \label{eqn:angular_func}
f_{\text{A}}(\theta) \equiv \sum_{i=0}B_{i} e^{-(\cos[\theta] - C_{i})^{2}/(2\sigma_{i}^2)}
\end{equation}
where $\{B_{i},\sigma_{i}\}$ is the set of flexible parameters with fixed $\{C_{i}\}$ sampled from the range [-1, 1] at 200 evenly spaced points. Independent parameter sets for pairs and triplets of species were employed for each of the five energy contributions (Eq.~\ref{eqn:two_body}-\ref{eqn:three_body_mf_2}). Further simplification was achieved by enforcing invariance to permutation of atoms in the two body terms of Eqn.~\ref{eqn:two_body} and to permuting the neighbor (non-central j and k) atoms in any angular function. For the wave-vectors, we used 400 evenly spaced wave-vectors in the range $(0,4\pi)$.

In an effort to regularize our ML potential, we added in a core repulsion of the Weeks-Chandler-Anderson (WCA) form at close atom-atom separations as well as a finite range cutoff between atoms. For a pair of atom types, the hard core was set to values slightly ($\sim0.2 \ \text{\AA}$) below the minimum inter-atomic distance between those atom types encountered in the training data. The cutoff distance was set to $r_{\text{C}}=3.0 \ \text{\AA}$. Before the hard cutoff, a smooth cutoff using the standard Sigmoid form, $S(r)$, with a switching distance of $\delta r=0.02 \ \text{\AA}$ is applied to all radial functions of the form $S((r_{\text{C}}-4\delta r - r_{i,j})/\delta r)$ to ensure negligibly small forces at the hard cutoff. An inner smooth cutoff is also applied to every radial function via $S((r_{i,j} - r_{\text{I}})/\delta r)$ where $r_{\text{I}}=0.3 \ \text{\AA}$.

Through an iterative process, we created an initial ML model that was periodically retrained to newly acquired simulation data. The initial data was obtained via Density Functional Theory Tight Binding simulations in CP2K using the UFF force field with UFF dispersion corrections. 1000 configurations evenly spaced in the density range of $0.7-2.5 \ \text{g/cc}$ at simulation temperatures of $4000\ \text{K}$, $6000\ \text{K}$, $8000\ \text{K}$ and $10000\ \text{K}$ were collected. Energy and forces were recalculated within the Kohn-Sham DFT formalism, outlined in Sect.~\ref{dft}, on the entire aggregated data-set of configurations using electronic temperatures of $1000\ \text{K}$, $4500\ \text{K}$, and $8000\ \text{K}$. The initial ML model was fit to this mixed electronic temperature data-set with the goal of realizing a model that performs well across all conditions. 
Retraining with additional DFT simulation configurations took place at unspecified intervals whenever the acceptance rate for NMC (using chain lengths of 30-40) was observed to drop below ~40\%. Retraining ceases upon approach to equilibrium.

Overall, our NMC approach provided a significant acceleration over an equivalent (non-nested) MC approach for a comparatively small upfront investment comprised of generating training data and training the ML models. The acceleration of our NMC scheme for the nested moves (swap and cluster), relative to bare MC, was $\mathcal{O}(10)$.~\cite{Gelb,HeAr,Iftimie,ShockNitrogen} Even with the acceleration of swap and cluster moves, the 225 simulations totaled $\mathcal{O}(1\text{M})$ CPU hours to reach the final set of equilibrium samples. The total cost of the simulations far outweighs the initial training data generation cost of $\mathcal{O}(10\text{K})$ CPU hours. Occasional training/retraining of the ML reference potential is even less burdensome; it was performed on a standard desktop in a few hours.

\section{Equation of state model}
\label{eos_model}

As mentioned in Sect.~\ref{eos}, we decompose the Helmholtz free energy per volume ($a$) into ideal ($a_{\text{id}}$) and excess ($a_{\text{ex}}$) contributions. Model flexibility is embedded in the excess free energy contribution
\begin{equation} \label{eq:helmholtz_excess}
a_{\text{ex}} \equiv n e_{0} + \sum_{m=2}^{O} \sum_{i=1}^{m-1} \lambda_{i} \Gamma_{n}^{i} \Gamma_{T}^{m-i}
\end{equation}
where $e_{0}$ is an optimized baseline reference energy per atom, $O$ is the order of the multi-variable polynomial expansion,
\begin{equation} \label{eq:density_term}
\Gamma_{n} \equiv \dfrac{(n/\overline{n})^{2}}{\delta_{n}+n/\overline{n}}
\end{equation}
and
\begin{equation} \label{eq:temperature_term}
\Gamma_{T} \equiv \dfrac{\delta_{T,1}+T/\overline{T}}{\delta_{T,2}+T/\overline{T}}
\end{equation}
and $n$ is the number of atoms per unit volume (number density) and $\lambda_{i}$ are optimized weights. The density and temperature shift factors ($\delta_{n},\delta_{T,1},\delta_{T,2}$) are also optimized while the reference density ($\overline{n}$) and temperature ($\overline{T}$) are set to the mean of the density and temperature in the data set.

The choice of the above form enforces a few key features. The first, is the limit to ideality with vanishing density. While a better limiting view at low density is an ideal gas of molecules, this is more challenging to capture in a fully reactive context where molecules are emergent entities. Second, the choice of $\Gamma_{n}$ provides a low density crossover from quadratic to linear with increasing density. This provides the correct low density quadratic correction on top of the reference energy (i.e., corrections must come from 2-body and higher order correlations) while capturing the empirically observed linear-like behavior at higher densities. Third, the form of $\Gamma_{T}$ yields a finite ``cold curve'' (ground state) contribution as $T\rightarrow 0$ as well as a finite contribution as $T\rightarrow \infty$, corresponding to an unweighted average of the potential energy over configuration space.

For a fixed set of $\{n, T\}$ points with corresponding ${e, P}$ measurements, we fit the model according to the loss function $\sum_{i}[\Delta_{e,i}^{2} + \Delta_{P,i}^2] / (2M)$ 
where $\Delta_{e}$ and $\Delta_{P}$ are the respective energy and pressure differences between the model and the DFT simulation data and $M$ is the number of measurements. 3-fold cross validation is used to identify the optimal model order ($O$) for training to the whole data set. Our model is implemented in the PyTorch machine learning package.~\cite{PyTorch}

For extracting uncertainties in all EOS derived quantities, we employ bootstrap re-sampling. Raw simulation derived ${e, P}$ data points are re-sampled with replacement to create 100 new data sets of the same size as the original. A cross-validated model is fit to each of the data-sets to yield an ensemble of EOS models, from which an ensemble of any derived parameter can be arrived at. From the ensemble we calculate the 50th percentile (median) to indicate the predicted value as well as the 2.5th (lower) and 97.5th (upper) percentile bounds to account for the 95\% confidence intervals.

\section{\texttt{Magpie} products}
\label{Magpie_products}

\setcounter{table}{0}
\renewcommand{\thetable}{D\arabic{table}}

\begin{table}[H]\centering
\begin{tabular}{ |p{3cm}||p{3cm}|  }
 \hline
 \multicolumn{2}{|c|}{\texttt{Magpie} products} \\
 \hline
 Common Name & Formula \\
 \hline
 atomic hydrogen & $\text{H}$\\
 atomic carbon & $\text{C}$\\
 atomic nitrogen & $\text{N}$\\
 atomic oxygen & $\text{O}$\\
 water & $\text{H}_{2}\text{O}$\\
 molecular hydrogen & $\text{H}_{2}$\\
 molecular nitrogen & $\text{N}_{2}$\\
 molecular oxygen & $\text{O}_{2}$\\
 carbon dioxide & $\text{CO}_{2}$\\
 carbon monoxide & $\text{CO}$\\
 nitric oxide & $\text{NO}$\\
 nitrogen dioxide & $\text{NO}_{2}$\\
 nitrous oxide & $\text{N}_{2}\text{O}$\\ 
 methane & $\text{CH}_{4}$\\
 ammonia & $\text{NH}_{3}$\\
 isocyanic acid & $\text{HNCO}$\\
 formic acid & $\text{CH}_{2}\text{O}_{2}$\\
 \hline
\end{tabular}
\caption{\label{tab:Magpie_products} Molecules used in the \texttt{Magpie} thermochemical calculation of the equilibrium product states. All species are uncharged within \texttt{Magpie}; however, the chemical formulas are more generally used in this article to denote a molecular aggregate with the bonding topology specified by the common name.}
\end{table}

\section{Molecular population model}
\label{mol_pop_model}

Molecular populations were modeled using a custom analytical form implemented in PyTorch~\cite{PyTorch}
\begin{equation} \label{eq:composition}
f_{\text{M}} \equiv \Bigg[\sum_{m=2}^{O} \sum_{i=1}^{m-1} \kappa_{i} \Lambda_{n}^{i} \Lambda_{T}^{m-i}\Bigg]^{2}
\end{equation}
where $f_{\text{M}}$ is the number of molecules of a given type yielded per molecule of high explosive (PETN) and
\begin{equation} \label{eq:composition_density_term}
\Lambda_{T} \equiv \dfrac{\Delta_{\rho,1}+\rho/\overline{\rho}}{\Delta_{\rho,2}+\rho/\overline{\rho}}
\end{equation}
and
\begin{equation} \label{eq:composition_temperature_term}
\Lambda_{T} \equiv \dfrac{\Delta_{T,1}+T/\overline{T}}{\Delta_{T,2}+T/\overline{T}}
\end{equation}
The density and temperature shift factors ($\Delta_{\rho,1},\Delta_{\rho,2},\Delta_{T,1},\Delta_{T,2}$) are all optimized along with the weights $\kappa_{i}$ while the reference density ($\overline{\rho}$) and temperature ($\overline{T}$) are set to the mean of the density and temperature in the data set. The squared form ensures the predictions are positive definite and the form of the density and temperature variables yield asymptotically finite contributions in the cold and low density limits. Models were fit according to the following loss function $\sum_{i}\Delta_{f_{\text{M}},i}^{2} / M$ 
where $\Delta_{f_{\text{M}}}$ is the population difference between the prediction and DFT simulation and $M$ is the number of measurements. Population uncertainties were estimated using the same protocol outlined in Sect.~\ref{eos_model} for the equation of state model.


%


 \end{document}